\documentclass[superscriptaddress]{revtex4}
\usepackage{amssymb}
\usepackage{amsmath}
\usepackage{graphicx}
\usepackage[normalem]{ulem}
\usepackage[dvips]{color}
\usepackage{array,longtable,multirow}
\usepackage{latexsym}
\usepackage{braket}
\begin{document}
\title{Effect of different filling tendencies on the spatial quantum Zeno effect}

\author{Xin Zhang}
\affiliation{Department of Physics, Nanjing University,
Nanjing 210008, China}

\author{Chang Xu \footnote{Corresponding author}\footnote{cxu$@$nju.edu.cn}}
\affiliation{Department of Physics, Nanjing University,
Nanjing 210008, China}

\author{Zhongzhou Ren \footnote{Corresponding author}\footnote{zren$@$nju.edu.cn, zren@tongji.edu.cn}}
\affiliation{Department of Physics, Nanjing University,
Nanjing 210008, China}
\affiliation{School of Physics Science and Engineering, Tongji University, Shanghai 200092, China}

\author{Jie Peng}
\affiliation{Laboratory for Quantum Engineering and Micro-Nano Energy Technology
and School of Physics and Optoelectronics, Xiangtan University, Hunan 411105,
People¡¯s Republic of China}

\begin{abstract}
The quantum Zeno effect is deeply related to the quantum
measurement process and thus studies of it may help shed
light on the hitherto mysterious measurement process in
quantum mechanics. Recently, the spatial quantum Zeno
effect is observed in a Bose-Einstein condensate depleted
by an electron beam. We theoretically investigate how different
intrinsic tendencies of filling affect the quantum Zeno effect
in this system by changing the impinging point of the electron
beam along the inhomogeneous condensate. Surprisingly, we find
no visible effect on the critical dissipation intensity at
which the quantum Zeno effect appear. Our finding shows the
recent capability of combining the Bose-Einstein condensate
with an electron beam offers a great opportunity for studying
the spatial quantum Zeno effect, and more generally the dynamics
of a quantum many-body system out of equilibrium.
\end{abstract}

\maketitle

\clearpage
\newpage

\section{Introduction}
Due to the peculiar laws of quantum mechanics, frequent enough
observation can halt the evolution of a system. This is known
as the quantum Zeno effect \cite{mis1977}. It is deeply connected to the
measurement process in quantum mechanics \cite{hom1997}. Studies of the quantum
Zeno effect may thus help shed light on the fascinating and
hitherto mysterious nature of the quantum measurement process \cite{leg2005,bru2017}.
The quantum Zeno effect has been realized experimentally in various
different physical systems \cite{ita1990,nag1997,fis2001,str2006,
bal2002,kwi1995,kwi1999,hos2006,ber2008,wol2013,Bar2013,zhu2014,pat2015}.
Interestingly however, most of the these
experiments are performed in quantum mechanical systems involving
several discrete energy levels, while the quantum Zeno effect of
the continuous spatial distribution of a quantum mechanical
system has been largely unexplored.
Recently, such spatial quantum Zeno effect has been observed in
a Bose-Einstein condensate \cite{Bar2013}, where a stronger electron beam depleting
the condensate counterintuitively led to less depletion.

From another angle of view, previous theoretical studies
\cite{fea1992,alt1994,gag1993,wal2001,por2014,por2016}
of the spatial quantum Zeno effect have been mainly concerned
with single-particle motion, while how such effect
occurs in a many-body system is less studied. The above system
using a Bose-Einstein condensate offers a great opportunity to
study such many-body spatial quantum Zeno effect.

Concerning this specific system, previous works \cite{Bar2013,bra2009,zez2012} have
predicted the existence of the quantum Zeno effect within this system,
and have studied how the characteristics of the time evolution depends
on the relation between the width of the depletion region and the
speed of sound in the condensate \cite{bra2009},
how the magnitude of the quantum Zeno effect depends on the width of the depletion
and the strength of interatomic interaction\cite{Bar2013},
how the existence of the quantum Zeno
effect depends on  how fast the depletion intensity decays in its wings,
as well as the existence of different steady-flow
patterns under depletion \cite{zez2012}.

In the present work, we study using
a time-dependent approach how the spatial quantum Zeno effect in this
system depends on the intrinsic tendency of the system to fill in the
depletion region.

Intuitively, if the electron beam impinges at places
where the trapping potential is higher and the density is lower, for example at the wings of a
harmonic trap rather than at the center, the intrinsic tendency of the system to fill in these
places will be higher.
From another point of view, the kinetic energies of the atoms
near such impinging position would be lower.
An interesting question can be raised about
how this affects the quantum Zeno effect.
One might naturally expect the spatial quantum Zeno effect to be more easily observed, i.e., to
appear at a lower dissipation strength
when the dissipation happens at a point with lower tendency of filling.
Surprisingly, our investigations
show that except for an overall scaling factor,
different impinging points do not affect the
spatial quantum Zeno effect, and in particular do not affect
the critical dissipation strength at which the quantum
Zeno effect starts to appear.

Our paper is organized as follows. In section II we describe the physical
system and the theoretical formalism used in our investigation.
In section III we give the numerical results and discussion.
In section IV we give our conclusion.

\section{Physical system and theoretical formalism}

\begin{figure*}[htb]
\centering
\includegraphics[width=14.0cm]{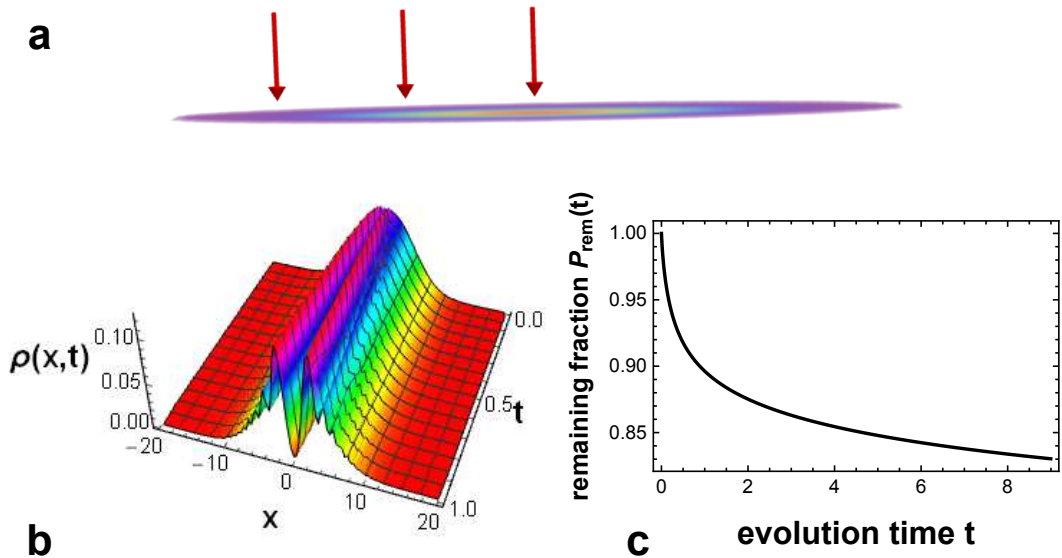}
\caption{(\textbf{a}) The physical system. A one-dimensional Bose-Einstein condensate
is trapped in a harmonic trap. An
electron beam (red arrows) knocking atoms out of the condensate can impinge at different
chosen positions of the condensate.
(\textbf{b}) The time evolution of the density distribution $\rho(x,t)$ of the condensate. The electron
beam burns a hole at its impinging point and depletes the condensate.
The time evolution is calculated using equation (\ref{tdGP}) and $\rho(x,t)=|\psi(x,t)|^2$.
(\textbf{c}) The remaining fraction
$P_{\textrm{rem}}(t)\equiv \int_{-\infty}^{\infty}  \rho(x,t) dx$
of the condensate as a function of time.
In (b) and (c), $w=0.1$, $x_d=0$, $\gamma=23.4$.
}\label{fig1}
\end{figure*}

A schematic depiction of the physical setup is shown in Fig. \ref{fig1}(a).
It constitutes of a one-dimensional Bose-Einstein condensate in a harmonic trap
under the action of an impinging electron beam knocking atoms out of
the condensate.
The impinging point of the electron beam can be changed along the one-dimensional condensate.
The evolution of the
system is governed
by the time-dependent Gross-Pitaevskii equation with a dissipation
term \cite{Bar2013,bra2009,zez2012}:
\begin{eqnarray}\label{tdGP}
i \frac{\partial \psi(x,t)}{\partial t}=- \frac{\hbar^2}{2 m}
\frac{\partial^2}{\partial x^2}\psi(x,t)+ g |\psi(x,t)|^2 \psi(x,t)
+V(x) \psi(x,t)-i \Gamma(x) \psi(x,t).
\end{eqnarray}
Here $\psi(x,t)$ is the wave function of the Bose-Einstein condensate,
$g$ is the nonlinearity parameter arising from atom-atom interaction.
In this work we have used $g=0.1$.
$V(x)= v_h x^2$ is the harmonic trapping potential.
Without loss of generality, we have set $m=\hbar=1$, $v_h=0.0005$, and normalized
the initial $\psi(x,0)$
such that $\int_{-\infty}^{\infty} |\psi(x,0)|^2=1$.
$\Gamma(x)=\gamma e^{-(\frac{x-x_d}{w})^2}$ describes
the dissipation due to the impinging electron beam.
$\gamma$ characterizes the intensity of the electron beam and thus the intensity
of the dissipation,
$w$ characterizes the width of the electron beam while $x_d$ is the
coordinate of the center of the beam.

The initial wave function is got by solving
for the ground state of the time-independent Gross-Pitaevskii
equation without dissipation:
\begin{eqnarray}\label{tidGP}
\mu \psi_0=- \frac{\hbar^2}{2 m}
\frac{d^2}{d x^2}\psi_0+ g |\psi_0|^2 \psi_0 +V(x) \psi_0,
\end{eqnarray}
where $\mu$ is the chemical potential.

In the numerical calculations, we first find the ground state
of equation (\ref{tidGP}) using the optimal damping algorithm \cite{dio2007}.
Then using this as the initial condition, we time-integrate
equation (\ref{tdGP}) to get the time evolution.

\section{numerical results and discussions}

An exemplary time-evolution of the density distribution of the condensate is shown
in Fig. \ref{fig1}(b). Initially, the condensate is in the ground state. The electron
beam acts a sink for atoms and burns a hole at the impinging position.
Due to the depletion from the electron beam, the remaining fraction of the condensate is a decreasing
function of time, as shown in Fig. \ref{fig1}(c).

\begin{figure*}[htb]
\centering
\includegraphics[width=12.0cm]{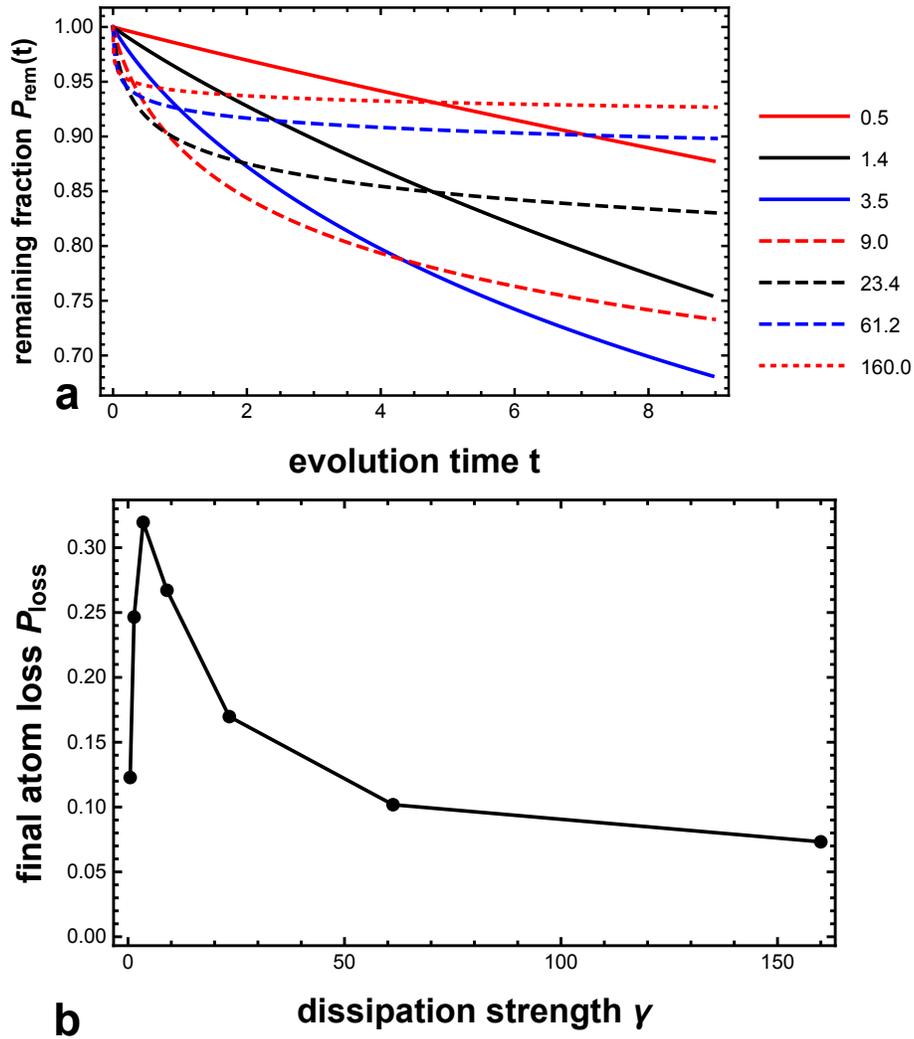}
\caption{(\textbf{a}) The remaining fraction of the condensate $P_{\textrm{rem}}(t)$ as a function of time,
for different intensities of the electron beam.
The spatial quantum Zeno effect occurs: starting
from the intensity parameter $\gamma=9.0$, as the dissipation
further increases, the final remaining fraction increases. (\textbf{b})
The remaining fraction at the end of evolution $P_{\textrm{loss}}$ as a function of the dissipation-intensity
parameter $\gamma$. The quantum Zeno effect is manifested as the decreasing part of the curve.
$w=0.1$, $x_d=0$.
}\label{fig2}
\end{figure*}

The spatial quantum Zeno effect can be
observed in this system. For example (Fig. \ref{fig2}(a)),
for the width parameter $w=0.1$, as the intensity
of electron beam becomes
larger, the total loss initially rises, but
then decreases as the intensity of dissipation
increases further. This is the quantum Zeno effect.
This can be seen more simply by plotting the total loss at the end
of evolution as a function of the dissipation strength parameter,
namely the intensity parameter of the electron beam $\gamma$  (Fig. \ref{fig2}(b)).
The existence of the decreasing part of this curve is the quantum Zeno effect.

\begin{figure*}[htb]
\centering
\includegraphics[width=16.0cm]{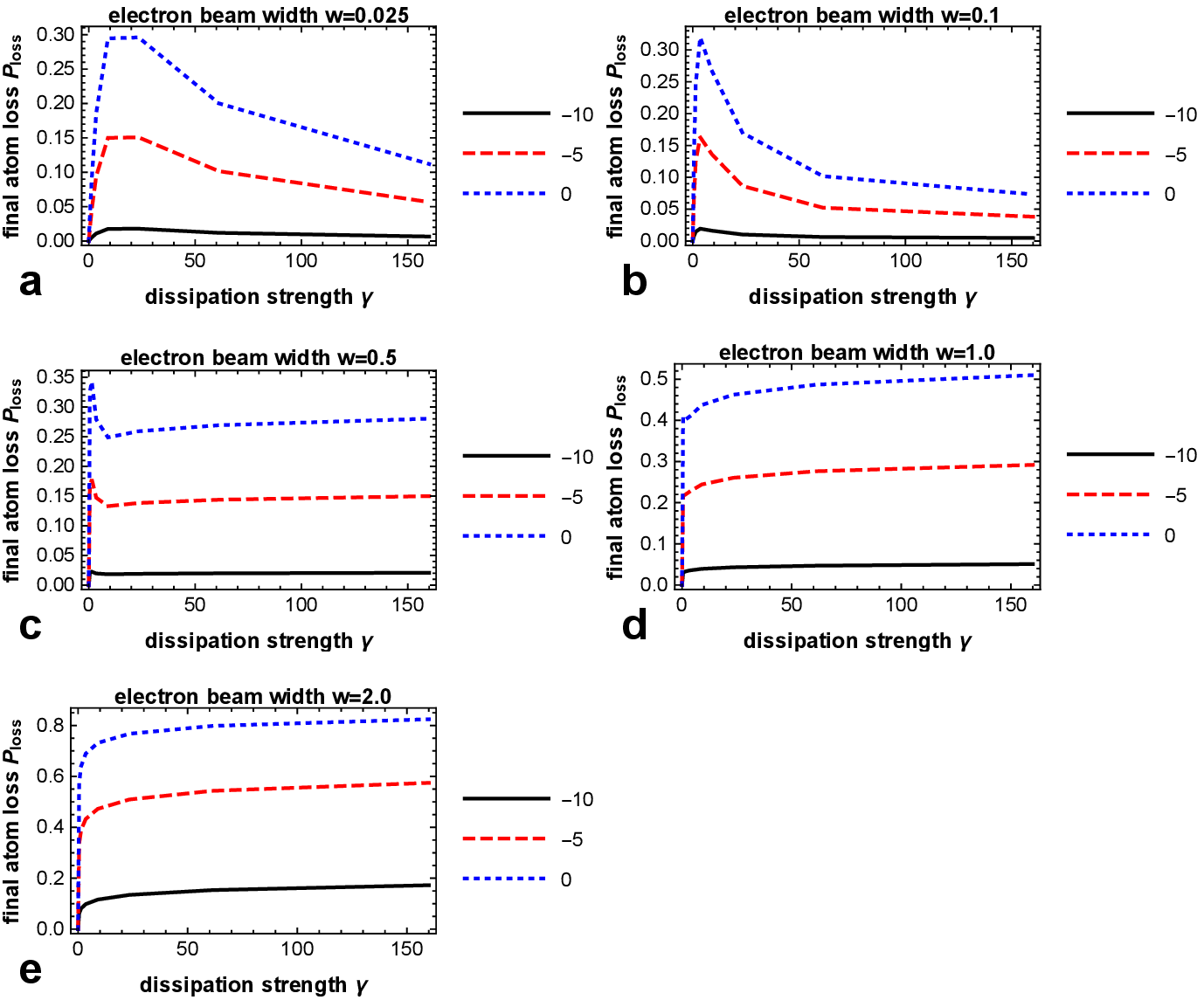}
\caption{
(a) The remaining fraction of the condensate $P_{\textrm{loss}}$ as a function of
the dissipation strength parameter $\gamma$, for three different impinging points
of the electron beam: at the center of the condensate ($x_d=0$), at the wing ($x_d=-5$),
and further out in the wing ($x_d=-10$). The width parameter $w=0.025$.
Different impinging points have
no visible effect on the starting value of $\gamma$ for the quantum Zeno effect,
but only adds an overall scaling factor to the curve. (b)-(e) Same as
(a), but for another four different widths $w=0.1, 0.5, 1.0, 2.0$ of the
electron beam. Different beam width affects the shapes of the curves significantly.
Remarkably, for each of these different beam width,
different impinging point always simply rescales the curve but has no visible effect
on the critical value of the dissipation strength parameter at which quantum Zeno
effect starts to appear.
}\label{fig3}
\end{figure*}

To investigate
how will the spatial quantum Zeno effect be affected by different
intrinsic tendency of filling,
we run the simulation with the electron beam shooting at different
positions of the inhomogeneous condensate. The results are given in Fig. \ref{fig3}.

In Fig. \ref{fig3}(a), the total loss as a function of the dissipation intensity
are shown for three different impinging points, represented by the black solid,
red dashed and blue dotted curves respectively. Surprisingly, while changing
the impinging point of the electron beam affects the absolute magnitude of the atom loss,
it has no visible effect on the critical dissipation intensity at which
the spatial quantum Zeno effect starts to show up. In fact, the shape of the curves for
different impinging points are very similar to each other except for an overall scaling factor.

In Figs. \ref{fig3}(b)-(e), the same simulations are run for another four different width parameters
$w$ of the electron beam.
In accordance with previous works, the shape of the curves are affected significantly by different
width of the electron beam. For a narrow electron beam, the quantum Zeno effect is more conspicuous (a).
For wider electron beams, the critical Zeno intensity field shifts to lower value (b)
and for even wider beams (d-e), the loss becomes a monotonically rising function
of the intensity of dissipation
and the quantum
Zeno effect is no longer visible.
Remarkably, for all the widths investigated corresponding to Figs. \ref{fig3}(a)-(e),
different impinging points have no visible effect on the critical Zeno intensity or the shape
of the curves, but only scales the atom losses by an overall factor.

\section{Conclusion}

In this paper we investigate the
effect of different intrinsic tendency of filling on the spatial
quantum Zeno effect by varying the impinging point
of the electron beam along the inhomogeneous
Bose-Einstein condensate in the simulations.
Surprisingly, impinging at different densities
do not change the critical dissipation intensity
at which the quantum Zeno effect starts to appear, but just adds an overall
scaling factor to the atom loss as a function of the dissipation strength. Our
finding shows the recent capability of
combining the Bose-Einstein condensate with an electron beam
offers a great opportunity for studying the spatial quantum Zeno effect, and
the dynamics of a quantum many-body system out of equilibrium.

\maketitle

\clearpage
\newpage

\,

\,

\,

\,

\begin{acknowledgments}
The work is supported by the National Natural Science
Foundation of China (Grant No. 11575082, No. 11761161001,
No. 11535004, No. 11375086, and No. 11120101005, No. 11235001) and by
the International Science \& Technology Cooperation Program
of China (Grant No. 2016YFE0129300).
\end{acknowledgments}

%\section*{Author contributions}
%X.Z. performed the calculations, X.Z., C.X., Z.R. and J.P. discussed the results and wrote the paper.

%\textbf{Competing financial interests:} The authors declare no competing financial interests.

\end{document}